# Students' accommodation allocation: A Multicriteria Decision Support System

Rôlin Gabriel Rasoanaivo [1][0000-0002-2496-2672] and Pascale Zaraté [1][0000-0002-5188-1616]

[1] IRIT, Université de Toulouse, CNRS, INP, UT3, UT1, UT2, Toulouse, France
`rolin-gabriel.rasoanaivo@ut-capitole.fr`,
`Pascale.Zarate@ut-capitole.fr`

**Abstract.** The social life of students at university has an impact on their educational success. The allocation of accommodation is part of this aspect. This article presents our proposal to improve students' allocation accommodation. We aim to support university administrative departments for the selection of students for housing.

Therefore, we propose a decision support system based on multi-criteria decision support methods. To calculate the weights of the criteria, we use the AHP method. Then, to rank the students, AHP, Weighted Sum Method and PROMETHEE methods are used.

The aim is to find the most adequate method to rank the students. The result is achieved because the AHP is able to calculate the weight of criteria and the AHP, SWM and PROMETHEE are able to rank the students.

## 1    INTRODUCTION

To successful manage universities, the institution must demonstrate its priorities in providing students housing because sharing accommodation and participating in campus life are part of the social and intellectual benefits of students [1]. According to researches carried out by Owolabi, on-campus students are more successful in their studies than off-campus students [2]. Therefore, following our research, several authors have been interested in the subject of student housing. Several studies collected the opinion of the students about their accommodation with the aim of bringing a new era of living on campus [3] [4] [5] [6] [7] [8] [9] [10]. Then other researches made the proposal of work tools to help the person in charge of housing within the universities, as well as the students to facilitate the search for housing [11] [12] [13] [14]. These works justify the importance of considering the management of student accommodation within university life.

However, each university has its own procedures for allocating student accommodation. The most common is the consideration of criteria which vary in number according to each institution. Thus, we have identified thirteen criteria for the allocation of accommodation  used by universities, namely: dependent children of parents, students with scholarships, distance from the place of study, admission to the examination, educational enrolment, age, nationality, level of study, date of application for accommodation, physical capacity of the student, orphan of a parent, parent working at the university, and year of undergraduate [15] [16] [17] [18] [19] [20] [21].

Our research has focused on Decision Support System (DSS) for students accommodation allocation. Let us note that the first DSS appeared in the 1960s [22]. Since then, several authors have investigated the application of these systems such as Eom et al. from 1971 to 2001 [23] [24] [25] [26] and Papathanasiou et al. from 1989 to 2019 [27]. Authors like Carlsson and Walden [28] have noted that Decision Support Systems are currently innovative and among the technological challenges. Keenan [29] has shown that they are spread over several areas including life sciences and biomedicine, physical sciences, and social sciences. Nevertheless, we did not find any DSS for students accommodation allocation.

Thus, our aim is to propose a decision support system based on AHP, WSM and PROMETHEE to facilitate the selection of student during the housing allocation process.

## 2    ISSUES

Regarding the criteria for housing allocation, each student is considered as a specific case. The idea is to select those students who meet them. However, there is much more demands than offers, but sometimes there are many more students who meet the criteria than there is housing to be allocated [17]. So, the question which remains is: what housing applications should be admitted or rejected? And which of these students will be housing allocated?

To do this, a method needs to be designed to fairly distribute this accommodation.

## 3    METHODOLOGY

Multi-criteria analysis methods are used to formulate real problems, according to three basics formulations: the choice problem, noted $P_\alpha$, the sorting or allocation problem noted $P_\beta$ and the ranking problem noted $P_\gamma$ [30]. The allocation of accommodation to students belongs to the ranking problem ($P_\gamma$). This situation led us to choose the following three methods: AHP, WSM and PROMETHEE to solve this problem. These three methods are selected because of their popularity and their usefulness.

### 3.1    Analytic Hierarchy Process (AHP)

AHP was developed by Saaty in 1970 [31]. According to its founder [32], the method is based on three concepts: hierarchical structure, priority structure and logical consistency. To find the logical consistency, the following calculations have to be made:

Medium consistency: $\lambda_{max} = a_{ij}\frac{w_j}{w_I}$     (1)

Consistency index: $CI = \frac{\lambda_{max} - n}{n-1}$          (2)

Consistency ratio: $CR = \frac{CI}{RI}$          (3)

Finally, when comparing pairs, the consistency ratio (CR) must be within 0.1. Otherwise, the results could be inconsistent.

### 3.2    Weight Sum Method (WSM)

The weighted sum method combines all criteria into one scalar composite objective function using the weighted sum [33]. The steps to follow are normalisation of all alternatives, normalisation of the weights whose sum must be equal to 1 and implementation of the weighted sum. The result is obtained by [34]:

$R(a_{ij}) = \sum_{j=1}^{n} w_j a_{ij}$          (4)

### 3.3    Preference Ranking Organisation METHods for Enrichement Evaluation

Brans initiated the PROMETHEE method in 1982 [35]. It is a multi-criteria method for defining the relationships of outranking, indifference, and incomparability between alternatives. Two concepts are to be considered, namely the preference index and the outranking flows. The method of calculating these flows [36] is presented below. The result is obtained by comparing the outflows, inflows, and net flows of the alternatives.

Preference index: $\begin{cases} \pi(a_1, a_2) = \sum_{j=1}^{k} P_j(a_1, a_2) w_j \\ \pi(a_2, a_1) = \sum_{j=1}^{k} P_j(a_2, a_1) w_j \end{cases}$          (5)

Outflow: $\Phi^+(a) = \frac{1}{n-1} \sum_{x \in A} \pi(a, x)$     (6)

Inflow: $\Phi^-(a) = \frac{1}{n-1} \sum_{x \in A} \pi(x, a)$          (7)

Net flow: $\Phi(a) = \Phi^+(a) - \Phi^-(a)$     (8)

## 4    PROTOTYPE

Let us recall that a DSS is structured by three main components which are: the model management system, the user interface, and the knowledge base [37]. We propose a DSS prototype based on the architecture defined by Sprague [38]. The DSS will be composed by a Data Base, a Model Base, and a Human/Interface module.

### 4.1    Database

The developed DSS consists of allocating housing to students through their applications. We have implemented a relational database management system using three methods AHP, WSM and PROMETHEE. We present in the following figures some
structure of this database.

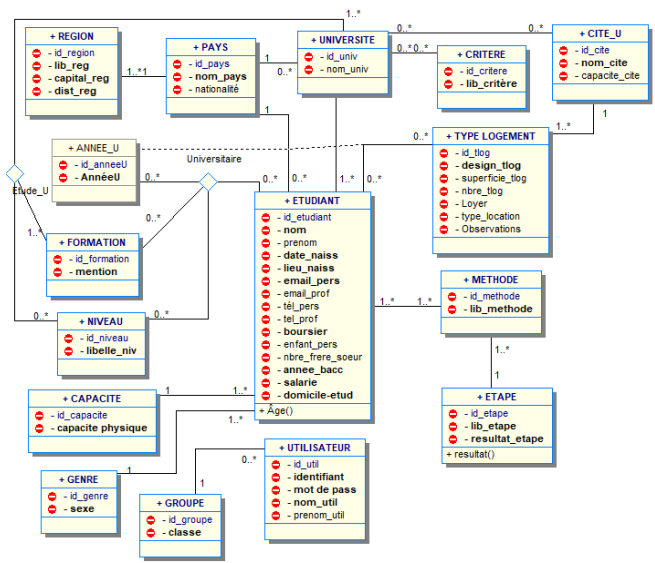

**Fig. 1.** Class diagram

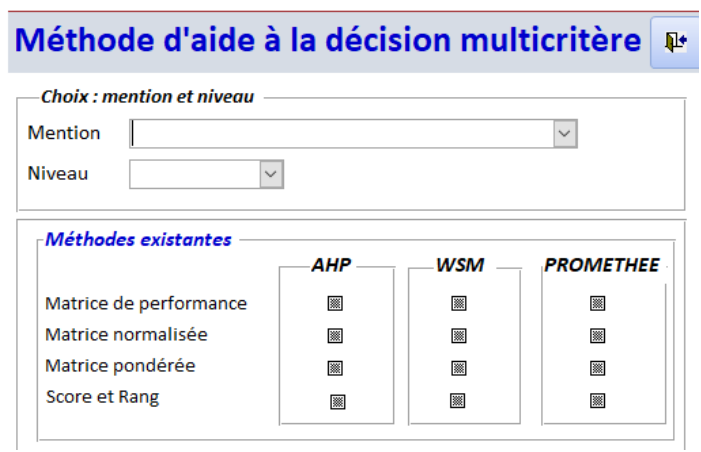

**Fig. 2.**  Choice of method

## 4.2    Process

The application concerns a university that has eleven criteria for evaluating students
[17]. These criteria are grouped into two sets, of which the first six are basic criteria

and the last five are social criteria. The following table 1 describes these evaluation criteria.

**Table 1.** Criteria for evaluating students at a university

| Basic criteria | Admission requirements | | Social criteria | Value |
|---|---|---|---|---|
| Age | By level | | Physical capacity (CP) | Normal = 5 ; Disability = 10 |
| Year of Baccalaureate | By academic year | | Orphan of parent (OP) | None = 0; Father or Mother = 5 ; Father and Mother = 10 |
| Administrative registration | Enrolled | | Parent's place of work (LTP) | University = 5 ; Other = 0 |
| Examination result | Successful | | Dependent child of parent (EC) | By number |
| Nationality | According to the case | | Distance from home (DD) | By mileage |
| Professional situation | Not employed | | | |

In relation to these two groups of criteria, the assessment procedure proceeds in two stages. Firstly, students must meet the basic criteria, otherwise their applications for housing will be rejected. Then, those who pass the basic criteria will move on to the second assessment where the social criteria are applied. Each social criterion has its own value to rank the students. This is where the multi-criteria decision support method comes in. At this stage, the processing is done by specialty and by level (bachelor/master/PhD). The following figure 3 summarizes the procedure for evaluating students by criteria.

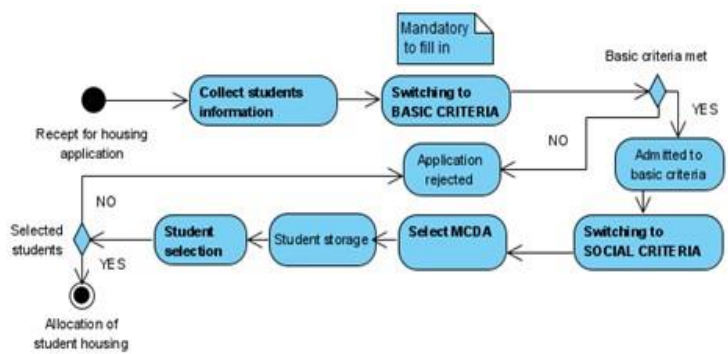

**Fig. 3.** processing applications for student accommodation

### 4.3   Result

**Assessment according to the basic criteria**

This concerns the housing applications of students in the first year of the Computer Science and Law degrees. The result is shown in Table 2 below.

**Table 2.** Result of the processing applications for student accommodation

| Mention | Application received | Student meets basic criteria | Application rejected |
|---|---|---|---|
| Computer science | 35 | 26 | 9 |
| Law | 101 | 78 | 23 |

The following figure 4 and figure 5 shows the extracts of students in the first year of the Computer Science and Law degrees admitted to the basic criteria.

**Fig. 4.** : Extract of L1 students in Computer Science admitted to the basic criteria

**Fig. 5.** Extract of L1 Law students admitted to the basic criteria

The figures 6 and 7 below show the extract of students who did not meet some basic criteria, and as a result their applications were rejected.

**Fig. 6.** Extract of L1 students in Computer Science does not meet some basic criteria

**Fig. 7.** Extract of L1 students in Law does not meet some basic criteria

**Assessment according to social criteria**

This second phase will deal with the 26 and 78 students meeting the basic criteria. We show for each of the three methods chosen, the result for the 26 students of the Computer Science major. To begin with, the initial judgement matrix and the judgement matrix normalized to scale 10.

**Fig. 8.** Initial judgement matrix

**Fig. 9.** normalised judgement matrix

*Result according to AHP*

The application of the comparison scale according to Saaty [40], allowed us to obtain the criteria judgement matrix and the result of the priorities presented in figure 10. Thus, the values of the elements of logical consistency are given below.

**Fig. 10.** Priority of criteria

Subsequently, establish the student judgement matrices for each criterion from the normalized judgement matrix. After various calculations, the students' priorities for each criterion are shown in Figure 11 and the ranking of students in Figure 12 below.

**Fig. 11.** priority of student

**Fig. 12.** AHP student ranking

*Result according to SWM*

The weights of the criteria considered are those calculated according to AHP. Figure 13 shows the application of the weighted sum to the normalized judgement matrix and Figure 14 gives the final result.

**Fig. 13.** implementing the weighted sum

**Fig. 14.** WSM student ranking

*Result according to PROMETHEE*

For the preferences: all the criteria are to be maximised, the weights applied are always those calculated according to AHP and the preference function for all the criteria are of the same Usual type. After setting the preferences and considering the assessments, the result will be given in figure 15.

| F i l u x | | | | | | | | | | | | | | | | | | | | | | | | | |
|---|---|---|---|---|---|---|---|---|---|---|---|---|---|---|---|---|---|---|---|---|---|---|---|---|---|
| Phi | 0.370 | 0.260 | 0.241 | 0.238 | 0.124 | 0.107 | 0.082 | 0.082 | 0.082 | 0.046 | 0.046 | 0.046 | -0.037 | -0.052 | -0.052 | -0.073 | -0.075 | -0.080 | -0.111 | -0.143 | -0.143 | -0.182 | -0.182 | -0.182 | -0.182 | -0.230 |
| Phi+ | 0.422 | 0.334 | 0.328 | 0.326 | 0.248 | 0.204 | 0.172 | 0.172 | 0.172 | 0.152 | 0.152 | 0.157 | 0.095 | 0.095 | 0.093 | 0.079 | 0.135 | 0.059 | 0.048 | 0.048 | 0.020 | 0.020 | 0.020 | 0.020 | 0.020 | 0.000 |
| Phi- | 0.052 | 0.074 | 0.087 | 0.088 | 0.124 | 0.090 | 0.090 | 0.090 | 0.090 | 0.106 | 0.106 | 0.106 | 0.194 | 0.147 | 0.147 | 0.166 | 0.155 | 0.215 | 0.170 | 0.190 | 0.190 | 0.202 | 0.202 | 0.202 | 0.202 | 0.230 |
| Rank | 1 | 2 | 3 | 4 | 5 | 6 | 7 | 7 | 7 | 10 | 10 | 10 | 13 | 14 | 14 | 16 | 17 | 18 | 19 | 20 | 20 | 22 | 22 | 22 | 22 | 26 |

**Fig. 15.** PROMETHEE student ranking

## 4.4    Discussion

In 2018, Saare et al. were designing a mobile system for managing and mitigating the accommodation problems at the Universiti Utara Malaysia [12]. In 2019, Podunavac et al. proposed a web portal managing registration for student accommodation in a dormitory [13]. In 2020, Magambo et al. have created an online portal for locating Students' private rental Accommodation in Tanzania [14].

For our part, we developed a DSS focusing on student housing allocation using three multi-criteria decision support methods (AHP, WSM and PROMETHEE) to better compare the results obtained. And to make a choice between the method used by the university, we illustrate in the following figure 16 the comparison of the results obtained and in terms of number percentage in the table 3 the similarity of student ranks.

| | | | | | | | | | | | | | | | | | | | | | | | | | | |
|---|---|---|---|---|---|---|---|---|---|---|---|---|---|---|---|---|---|---|---|---|---|---|---|---|---|---|
| AHP | | 1 | 3 | 2 | 4 | 9 | 5 | 6 | 6 | 10 | 10 | 15 | 13 | 16 | 17 | 19 | 18 | 19 | 19 | 21 | 21 | 21 | 26 | | | |
| WSM | Rang | 1 | 3 | 2 | 4 | 12 | 5 | 6 | 6 | 9 | 9 | 18 | 14 | 14 | 16 | 25 | 17 | 19 | 19 | 21 | 21 | 21 | 26 | | | |
| PROMETHEE | | 1 | 2 | 3 | 4 | 5 | 6 | 7 | 7 | 10 | 10 | 13 | 14 | 14 | 16 | 17 | 18 | 19 | 20 | 20 | 22 | 22 | 22 | 22 | 26 | |

**Fig. 16.** comparison of results

From the comparison of these three results, we can draw the following three situations:

- 7.69% of students have the same rank on the three methods: two students;
- 15.38% have different ranks on the three methods: four students;
- 76.92% have the same rank on two methods: the remaining twenty.

To make a choice between the method used by the university, we illustrate the two-by-two comparison of the results obtained.

**Table 3.** similarity of student ranks on two methods

| | WSM | | PROMETHEE | |
|---|---|---|---|---|
| | Number | % | Number | % |
| **AHP** | 8 | 30.77 | 13 | 50 |
| **WSM** | | | 4 | 15.38 |

## 5 CONCLUSION AND PERSPECTIVE

To decide is certainly to take risks but it is to be in reaction to a strategic choice to be made. The same applies to the person in charge of university work when receiving requests for student accommodation, a decision must be made whether to accept or reject an application. But, in any case, the question must always be asked: why was it accepted or rejected?

This is the reason for our analysis, which proposes the use of multi-criteria decision-making methods to classify students in relation to their situation, regarding the criteria for allocating housing. Three methods were studied: AHP, WSM and PROMETHEE.

The result showed us that each method was able to rank the students. And after comparing the results, we found that the rankings of the students are different for each method. First, the ranking carried out by AHP showed us that compared to WSM the similarity of ranks is 30.77%, and compared to PROMETHEE, it becomes 50%. Secondly, as for WSM, this similarity of ranks with PROMETHEE is 15.38%.

However, since the purpose of the decision analysis is to clarify the choice of the decision maker and not to replace the decision maker, the manager of university services has the choice of the method to use according to the analysis of the results that we carried out.

Nevertheless, we can report that a difficulty was encountered when using the AHP method. This being when the criteria or the alternatives are more numerous. For example, in the case of the 78 students in the Law stream and the other applications accepted in the other streams, the students are even more numerous. It will be difficult to handle such a large square matrix.

So, in our future research, we will use the AHP method only to calculate the criteria weights. On the other hand, WSM and PROMETHEE for ranking students. And in addition, we will look for another method that is easier to handle and combine with these two methods for storing students. Then, we will not stop with a simple application of the methods, but we will look for a trick allowing to recover all the ranks of the students carried out by the methods implemented in the system and to reorder the students from all their ranks.

Finally, our objective in this first analysis is then to integrate all these ranking methodologies into the developed prototype, offering end users the possibility of having several methodologies to use. And in the perspective, it will be to find the possibility of using all the results of these methods to obtain a new ranking of the students.